
\font\twelvebf=cmbx12
\font\ninerm=cmr9
\nopagenumbers
\magnification =\magstep 1
\overfullrule=0pt
\baselineskip=18pt
\line{\hfil CCNY-HEP 2/94}
\line{\hfil March 1994}
\vskip .8in
\centerline{\twelvebf Hamiltonian Analysis of the Effective Action for
Hard}
\centerline{\twelvebf Thermal Loops in QCD}
\vskip .5in
\centerline{\ninerm V.P. NAIR}
\vskip .1in
\centerline{ Physics Department}
\centerline{City College of the City University of New York}
\centerline{New York, New York 10031.}
\vskip 1in
\baselineskip=16pt
\centerline{\bf Abstract}
\vskip .1in
The effective action for hard thermal loops in Quantum
Chromodynamics (QCD) is related to a gauged Wess-Zumino-Novikov-Witten
theory.
Some of the technical issues of this approach are clarified and the
Hamiltonian formulation is presented. The two-point correlator
for the induced current in QCD is obtained; some simplifications
of the dynamics of the longitudinal modes are also pointed out.
\vfill\eject
\footline={\hss\tenrm\folio\hss}
\def \half {{\textstyle {1\over 2}}}
\def \vx {{\vec x}}
\def \vy {{\vec y}}
\def \twthird {{\textstyle {2\over 3}}}
\def \vp {{\vec p}}
\def \vQ {{\vec Q}}
\def \12 {{\textstyle {1\over 2}}}
\def \vx {{\vec x}}
\magnification =\magstep 1
\overfullrule=0pt
\baselineskip=18pt
\pageno=2
\noindent{\bf I. Introduction}
\vskip .2in

The possibility of obtaining the quark-gluon plasma phase of hadronic
matter in heavy ion collisions or perhaps in certain astrophysical contexts
provides an adequate physical motivation for studying Quantum
Chromodynamics (QCD) at high temperatures.  In a perturbative analysis,
hard thermal loops have to be calculated in order to carry out the
resummation or rearrangement of thermal perturbation theory necessary
to achieve a consistent expansion in powers of the coupling constant.
Hard thermal loops
 are also part of the effective action for the long wavelength
excitations of the plasma. They have therefore been
the focus of many recent investigations [1-6].

The generating functional $\Gamma [A]$ for the hard thermal loops is a
gauge-invariant nonlocal functional of the gauge potential $A_\mu$ and
is essentially an electric mass term for the gluons.
The current given by
$\Gamma [A]$  leads to a gauge-invariant description of Debye screening
and Landau damping effects [3].
$\Gamma [A]$ also has an elegant
mathematical structure, being closely related to the eikonal for a
Chern-Simons theory [2]. The equations of motion, incorporating
contributions from $\Gamma [A]$ are
nonlocal, but it
is possible to rewrite them as local, but coupled, equations [3].
Further, one can introduce an auxiliary field $G$,
which has values in the color group, and an action $\Gamma [A,G]$
which
gives the local equations of motion [5]. The elimination of $G$ leads back to
the
nonlocal functional $\Gamma [A]$.  $~\Gamma [A,G]$ is related to the
gauged Wess-Zumino-Novikov-Witten (WZNW) theory. A useful result
which emerged in a rather simple way within this formulation is the
positivity of the Hamiltonian in the hard thermal loop approximation.
Actually, one can go further. The local nature of
$\Gamma [A,G]$ makes it a convenient
starting point for the inclusion of hard thermal loop effects in the
systematic calculation of any physical quantity. With this in mind,
in this paper, we shall carry out the Hamiltonian analysis for this effective
action. This will lead to a current algebraic description of hard thermal
loops. A simple framework for the computation of current correlations
will also emerge from this analysis.
The two-point correlation function
for the induced currents in the hard thermal loop
approximation, which is related to the gluonic structure functions of the
plasma, is obtained to the lowest order. We also point out some simplifications
for the dynamics of the longitudinal excitations.

In section II, we give a brief resume of the effective action, set up
the Hamiltonian analysis and obtain the operator algebra.
In section III, the framework for
computing the correlators and the simplifications for the longitudinal modes
 are discussed. There are three appendices; the first discusses the issue of
whether
the coefficient of the effective action should be quantized, the second with
the
definition of the Hamiltonian and the third with canonical derivation of the
operator
algebra.
\vskip .1in
\noindent{\bf II.  The Effective Action, Hamiltonian and Current Algebra}
\vskip .2in
The gauge potential $A_\mu = (-it^a) A^a_\mu$, where $t^a$ are
hermitian $N\times N$-matrices representing the generators of the Lie
algebra of the color group; we take the color group to be $SU(N)$.
In addition to the gauge
potential $A_\mu$, we need the auxiliary field $G (x, {\vQ})$. This field
is an $SU(N)$-matrix depending on spacetime coordinates $x^\mu$ and
also on a unit three-dimensional vector $\vQ$. Further, $G$ satisfies the
condition $G(x, -\vQ )~=~ G^{\dagger} (x,\vQ )$. We shall also use the
notation $Q^\mu = (1, \vQ ),~~Q'^\mu = (1, -\vQ )$, and $A_{+}= {\12 }
A\cdot Q,~~A_{-}= {\12 } A\cdot Q' $. The effective action can be written
as [5]
$$
{\cal S}~= \int  -{\textstyle {1\over 4}}F^2 ~+~k~ \Gamma [A,G] \eqno(1)
$$
$$
\eqalignno{ \Gamma [A,G]~= \int d\Omega~\Bigl[ \int d^2x^T S(G) ~&+~{1\over
\pi}\int
d^4x~{\rm Tr}\bigl( G^{-1}\partial_{-}G A_{+}~ \cr
& -~A_{-}\partial_{+}G
 G^{-1}~+~A_{+}G^{-1}A_{-}G -A_{+}A_{-} \bigr) \Bigr] &(2)\cr }
$$
where the integration indicated by $d\Omega$ is over all orientations of
the unit vector $\vQ$. $S(G)$ is the WZNW-action for $G$, given by [7]
$$
S(G)~=~-{1\over 2\pi}\int_{M^2} d^2x~ {\rm Tr}(\partial_{+}G
G^{-1}\partial_{-}G G^{-1} ) ~+~ {1\over 12\pi} \int_{M^3} {\rm Tr}(dG
G^{-1})^3 \eqno(3)
$$
The integration in the first term is over the lightcone $M^2$ defined by the
variables $x^+= x_\mu Q^\mu ,~x^- = x_\mu Q'^\mu $; the second term, the
so-called Wess-Zumino (WZ) term,
requires extension of the fields to include one more coordinate, say $s$, and
corresponding integration. We can take $M^3 =M^2\times [0,1]$ with fields at
$s=1$
corresponding
to spacetime. Different ways of extending the fields to $s\neq 1$ will
give the same physical results. Since $S(G)$ has integration over the
lightcone variables,
the integration for  $S(G)$ in equation (2) is over
the transverse coordinates $x^T,~x^T \cdot \vQ =0$. $\Gamma [A,G]$ is
gauge invariant and, if we ignore the dependence on the transverse
coordinates, is just the standard gauged WZNW-action. Finally the
constant $k$ in equation (1) is $(N+ \12 N_F) {T^2\over 6}$, where $N_F$
is the number of quark flavors and $T$ is the temperature. The coupling
constant $g$ is
not explicitly indicated; it can be recovered by the replacement $A_\mu
\rightarrow
gA_\mu$.

The WZNW-action $S(G)$ obeys the combination rule
$$
S(hG)~=~ S(h)~+~S(G)~-~ {1\over \pi} \int_{M^2}~ {\rm
Tr}(h^{-1}\partial_{-}h~\partial_{+}G~G^{-1})\eqno(4)
$$
Using this rule, the equations of motion can be easily derived as
$$
\eqalignno{\partial_{+} {\cal A}_{-}~-~\partial_{-} A_{+} ~+~ [A_{+}, {\cal
A}_{-}]~=& ~0 &(5)\cr
(D_\mu F^{\mu\nu})^a ~-~J^{\nu a}=& ~0 &(6)\cr}
$$
$$
\eqalignno{
J^{\nu a}&= -{k\over 2\pi} \int d\Omega~ {\rm Tr}\left\{ (-it^a) [({\cal
A}_{-} -A_{-})Q^\nu ~+~ ({\cal A}_{+}-A_{+})Q'^\nu ]\right\}\cr
{}~&=-{k\over 2\pi} \int d\Omega~ {\rm Tr}\left\{ (-it^a) [G^{-1}D_{-}G ~
Q^\nu~-~ D_{+}G ~G^{-1} Q'^\nu ] \right\} &(7) \cr}
$$
$D_\mu$ denotes the covariant derivative, $D_\mu^{ac} =\partial_\mu
\delta^{ac}+
f^{abc}A^b_\mu $  and ${\cal A}_{+}= G
A_{+} G^{-1} -\partial_{+}G~G^{-1}, ~~{\cal A}_{-} =G^{-1}A_{-} G + G^{-1}
\partial_{-}G$.

The current given by equation (7) agrees with what is obtained by
evaluation of hard thermal loops, once the solution for ${\cal A}_{\pm}$
from equation (5) is substituted in. The action (2) is thus an acceptable
effective action at least at the level of the equations of motion.
Actually one can make this a stronger statement. The equation of
motion for the field $G$ has no independent solutions.  This can be seen
as follows. We can parametrize the potentials $A_{\pm}$ in terms an
$SU(N)$-matrix $V(x,\vQ)$ as
$$
A_{+}=~ -\partial_{+} V~V^{-1},~~~~~~~~A_{-}=~ -\partial_{-}V'~V'^{-1}
\eqno(8)
$$
where $V' (x, \vQ)= V(x,-\vQ)$. The general solution to equation (5) is
then given by
$$
G(x,\vQ)~=~ V(x, -\vQ) B(x^{+},x^T, \vQ) C(x^-, x^T, \vQ) V^{-1}(x,\vQ
)\eqno(9)
$$
where  $C$ is an arbitrary $SU(N)$-matrix depending on the variables indicated
and $B$ is given by $C$ with $\vQ \rightarrow -\vQ$. The matrices $B,C$
represent the new or independent degrees of freedom for the field $G$.
Notice, however, that the parametrization (8) of the potentials has
redundant variables; a transformation $V\rightarrow V U (x^-, x^T,\vQ) $,
with a corresponding change in $V'$, leaves the potentials invariant.
Further, ordinary gauge transformations act on the matrix $V$ as $V
\rightarrow h(x)V$. Thus the physical subspace for the components
$A_{\pm}$ is given in terms of the matrices $V$ with the identifications
$$
V(x,\vQ) \sim h(x) V(x,\vQ) U(x^-,x^T,\vQ) \eqno(10)
$$
The gauge freedom of multiplying $V$'s by matrices which do not depend
on $x^+$, viz. $U$'s in (10), shows that the we can reduce $G$ to just
$V(x,\vQ) V^{-1}(x,\vQ)$. There are thus no new real dynamical degrees
of freedom in $G$. The action can be simplified for this $G$ to
$$
\Gamma [A,G]=~ -\int d\Omega ~d^2x^T ~S\bigl( V^{-1}(x,-\vQ ) V(x,\vQ )
\bigr) \eqno(11)
$$
This agrees with the expression obtained by Feynman graph evaluation of
the hard thermal loops.

The Hamiltonian for the effective action (2) was given in ref.[5]
as
$$
{\cal H}=~\int d^3x~\left\{ {{E^2+B^2}\over 2} ~+~{k\over 8\pi} \int
d\Omega~ {\rm Tr}\left[  (D_0G~ D_0G^{-1})+(\vQ \cdot {\vec D}G~\vQ
\cdot {\vec D}G^{-1} \right] -A^a_0 {\cal G}^a \right\} \eqno(12)
$$
where $F^a_{0i}=E^a_i,~ F^a_{ij}=\epsilon_{ijk}B_k^a$ and
$$
{\cal G}^a= ~({\vec D}\cdot {\vec E})^a ~+~ {k\over 2\pi}\int d\Omega~
{\rm Tr}\bigl[ (-it^a) (G^{-1} D_- G~-~ D_+ G~G^{-1})\bigr] \eqno(13)
$$

${\cal G}^a=0$ is the Gauss law of the theory; it is also the time
component of the equation of motion for the gauge field. Expression (12)
makes it clear that the Hamiltonian is positive for all configurations
which are physical, i.e. obey the Gauss law. (It may be worth recalling
that, even for the Maxwell theory, the canonical Hamiltonian is positive
only for physical configurations.)

The Hamiltonian analysis is most easily carried out, as in the usual
cases, in the gauge $A_0^a=0$. We start by defining the currents
$$
J_+ ~=~ {k\over 4\pi} D_+ G~G^{-1}~=~ (-it^a)J^a_+
$$
$$
J_-~=~ -{k\over 4\pi} G^{-1}D_-G~=~ (-it^a)J^a_- \eqno(14)
$$
By virtue of the property $G^{-1}(x,\vQ )= G(x, -\vQ )$, these are related
by $J_+ (x, -\vQ )= J_- (x, \vQ )$.  The Hamiltonian can be written in
terms of these currents as
$$
{\cal H}=~ \int d^3x~\left\{ {{E^2+B^2}\over 2} ~+{2\pi \over k} \int
d\Omega~ (J^a_+ J^a_+ ~+~ J^a_- J^a_- ) \right\} \eqno(15)
$$
We have chosen the $A_0^a=0$ gauge; Gauss law must henceforth be
imposed as a constraint. ( For fixed $\vQ$, the integrand of the
second term involving the square of $J_+$ and the square of
$J_-$ is the Sugawara form of the Hamiltonian, well-known for two-dimensional
current algebras.) The equations of motion in the $A_0^a=0$ gauge
are
$$
\eqalignno{
E^a_i~ &=\partial_0 A^a_i &(16a)\cr
\partial_0 E^a_i ~+~ \epsilon_{ijk} (D_j B_k)^a ~&=~ \int d\Omega ~Q_i
(J^a_+ ~-~J^a_- ) &(16b)\cr
(D_- J_+)^a ~&=~ -{k \over 8\pi } E^a_i Q_i &(16c)\cr }
$$
Equation (16a) is just the definition of the electric field; however, in a
Hamiltonian approach it is an equation of motion and we have displayed
it as such.

The commutation rules must be such that equations
(16) follow as the Heisenberg equations of motion for the Hamiltonian
(15). Knowing the current algebra of the WZNW-theory, we can make a
guess as to what the appropriate commutation rules are for our
problem and verify them by checking
that they lead to equations (16) starting from the Hamiltonian (15).
The commutation rules are then seen to be
$$
\eqalignno{
[~E^a_i (\vx ), A^b_j(\vx ' ) ~] ~&= -i\delta^{ab}\delta_{ij} \delta(\vx -\vx
')
&(17a)\cr
[~E^a_i(\vx ), J^b_{\pm} (\vx ')~]~&= {\pm}i {k\over 4\pi} Q_i \delta^{ab}
\delta(\vx -\vx') &(17b)\cr
[~J^a_{\pm} (\vx, \vQ), J^b_{\pm}(\vx' , \vQ' )~]~&= if^{abc}
J^c_{\pm}\delta(\vx -\vx') \delta(\vQ, \vQ')\cr
& ~~~~~{\mp}{k\over 4\pi}
Q_i(D_x)^{ab}_i \delta(\vx -\vx') \delta (\vQ, \vQ') &(17c)\cr
[~J^a_+ (\vx, \vQ), J^b_- (\vx', \vQ')~]~&=0 &(17d)\cr}
$$
All other commutators vanish. $\delta (\vQ, \vQ')$ stands for the
$\delta$-function on the two-dimensional sphere
corresponding to the unit vector $\vQ$, i.e. $\int d\Omega_{\vQ'}
{}~\delta (\vQ,\vQ') f(\vQ') =f(\vQ)$.

We have checked that these rules obey the
Jacobi identity. Since we are postulating the algebra, this is a necessary
check. Of course, commutation rules can be obtained from
the action by standard quantization procedures. We have verified that
equations (17) follow from such a canonical approach as well. The canonical
method
is somewhat involved.  The most efficient way is
to use the language of symplectic structures. This is sketched
in the appendix. Finally, the condition $J^a_+ (x -\vQ)~-~J^a_- (x,\vQ)=0$
has to be imposed as a constraint, just like the Gauss law.

We now turn to deriving an equation for the correlators of currents.
The parameter $k$ in equation (17c) controls the semiclassical expansion, large
$k$ being the classical limit [7]. For a semiclassical expansion, it is
convenient to
use the currents $I_{\pm}= {\sqrt{4\pi \over k}}J_{\pm}$.
Eventhough we are not doing a semiclassical expansion at
this stage, it is useful to consider correlators of $I_+$'s. ($I_-$'s are
related
by $\vQ \rightarrow \vQ'$). Thus introducing a source $\alpha^a(x,\vQ)$ for
$I_+^a(x,\vQ)$ and a source $\eta^a_i(x)$ for $A^a_i(x)$, we can define a
generating
functional for correlators
$$
Z[\eta, \alpha ]= \exp (W [\eta, \alpha]) = \langle T \exp [ i\int d^4x~\{(\int
d\Omega~
\alpha^a  I_+^a )+\eta^a_i A^a_i\} ] \rangle \eqno(18)
$$
$W[\eta,\alpha]$ generates the connected functions. The angular brackets denote
a thermal
average or a vacuum expectation value
depending on whether the thermal effects of the soft modes
(momenta $\sim gT$) are to be included or not. We now derive $Z [\eta, \alpha
]$ with
respect to $\alpha$ and take the time-derivative. As is well known, this leads
to
a term involving the commutators and a term involving the time-derivative of
the
current $I_+^a$. This can be simplified using the equations of motion and the
operator
algebra  (17). The result is
$$
\eqalignno{
(\partial_0 +Q_i{\cal D}_i ) \langle I^a(x,\vQ)\rangle &+~f^{abc}\left(
{\sqrt{4\pi\over k}} ~\alpha^b \langle I^c\rangle ~+~ Q_i
\langle A^b_i(x) I^c(x,\vQ)
\rangle \right)\cr
&+{\sqrt{k\over 4\pi}}Q_i \partial_0 \langle A^a_i\rangle  -(Q_i{\cal D}_i
\alpha)^a~=0
&(19)\cr}
$$
where
$$
\eqalignno{
\langle I^a(x,\vQ)\rangle &= ( -i {\delta \over {\delta \alpha^a(x,\vQ)}}
)~W,\cr
\langle A^b_i(x) I^c(y,\vQ)\rangle &= ( -i {\delta \over {\delta
\eta^b_i(x)}})~
( -i{\delta \over {\delta \alpha^c(y,\vQ)}})~ W,~~{\rm etc.}&(20)\cr}
$$
The sources are not yet set to zero in equations (19,20). Further,
$ {\cal D}_i^{ac} = \partial_i \delta^{ac} + f^{abc} \langle A^b_i\rangle $.
In a similar way,
we can derive an equation for the quantity $\langle A^a_I(x)\rangle$. This is
given by
$$
{\partial^2 \over {(\partial x^0)^2}} \langle A^a_i(x)\rangle ~+~ ({\cal D}_j
{\cal F}_{ij})^a
{}~+~\Delta^a_i~+ \left[ {\sqrt{k\over 4\pi}} \int d\Omega~ Q_i(\alpha^a
-2\langle I^a\rangle )
\right] ~-~\eta^a_i~=0
\eqno(21)
$$
where
$$
\eqalignno{
\Delta^a_i~= f^{abc}\biggl[ \partial_j \langle A^b_i(x) A^c_j(x)\rangle &+
\langle A^b_j(\partial_i A^c_j-
\partial_jA^c_i)\rangle
+f^{crs} \Bigl( \langle A^b_j(x) A^r_i(x)\rangle \langle
A^s_j(x)\rangle \cr
&+ \langle A^b_j(x)A^s_j(x)\rangle
\langle A^r_i(x)\rangle +\langle A^b_j(x)\rangle \langle
A^r_i(x)A^s_j(x)\rangle \cr
&+\langle A^b_j(x)A^r_i(x)A^s_j(x)\rangle \Bigr) \biggr]
&(22)\cr}
$$
and ${\cal F}_{ij}^a= \partial_i \langle A^a_j\rangle -\partial_j\langle A^a_i
\rangle +f^{abc}
\langle A^b_i\rangle \langle A^c_j\rangle $.

Any correlation function involving the fields can be calculated using
the equations (19-22). It is, of course, possible to write functional integrals
for
the correlation functions. However, since the currents of the WZNW-theory are
the important
quantities for calculating the correlators of the fields $A^a_i$, the present
approach of
obtaining differential equations using the Hamiltonian framework is somewhat
simpler.
Equations (19-22) are written without the renormalization
factors; eventually, each average of an
independent opertor product should have such a factor.

As an application of these equations, we calculate the correlator
$\langle I^a(x,\vQ) I^b(y,\vQ')\rangle $ to the lowest order (with no loops).
Differentiating the
equations (19,21) with respect to $\alpha^b(y,\vQ')$ and setting all sources to
zero,
(indicated by a subscript zero on the angular average signs), we get
$$
\eqalignno{
\bigl[ (\partial_0^2 -\partial_k \partial_k )\delta_{ij} & +\partial_i
\partial_j \bigr]
\langle A^a_j(x)I^b(y,\vQ')\rangle _0 \cr
&+{\sqrt{k\over 4\pi}} \biggl[-i Q'_i\delta^{ab}\delta (x-y)-
2\int d\Omega'' ~Q''_i \langle I^a(x,\vQ'')I^b(y,\vQ')\rangle_0 \biggr]~\approx
0
&(23) \cr}
$$
$$
\eqalignno{
(\partial_0~+ Q_i\partial_i )\langle I^a(x,\vQ)I^b(y,\vQ')\rangle_0 ~&+
{\sqrt{k\over 4\pi}}
\partial_0 \langle A^a_i(x)I^b(y,\vQ')\rangle_0\cr
& ~+i(Q_i\partial_i )\delta^{ab}\delta(x-y)
\delta(\vQ,\vQ')~\approx 0. &(24)\cr}
$$
We have dropped the higher loop terms, as emphasized by the approximate
equality.
Also we note that it is consistent to take $\langle I^a\rangle =0$.
Equation (23) can be solved by using the Green's function
$$
G_{ij}(x,y)~=~\int {d^4p\over (2\pi)^4} e^{-ip(x-y)} ~~ {i\over
p^2}(\delta_{ij}
-{p_i p_j\over p_0^2} ) \eqno(25)
$$
The integrand is singular as $p_0 \rightarrow 0$, indicating the residual gauge
invariance under time-independent transformations. Nevertheless we can use this
in the usual way with some
prescription for handling the singularity such as the $\alpha$-prescription
[9].
Using the solution of (23) in (24) we get the equation
$$
\int d^4z~d\Omega''~\left[ (\partial_0+ Q_i\partial_i
)\delta(x-z)\delta(\vQ,\vQ'')
{}~+{ik\over 2\pi} \partial_0 (Q_i G_{ij}(x,z)Q''_j) \right] \langle
I^a(z,\vQ'')I^b(y,\vQ')
\rangle
$$
$$
{}~~~~~~~~~~~~~~~~~~~~~~~~~= \delta^{ab}\left[
-iQ_i\partial_i \delta (x-y) \delta(\vQ,\vQ')~+ {k\over 2\pi}
\partial_0 (Q_iG_{ij}(x,y)Q'_j)\right]
\eqno(26)
$$
The solution to this equation is given by
$$
\langle I^a(x,\vQ) I^b(y,\vQ')\rangle~= \delta^{ab} \int {d^4p\over (2\pi)^4}
e^{-ip(x-y)} ~F(p,\vQ,\vQ')
\eqno(27)
$$
$$
F(p,\vQ,\vQ')~= f_1 \delta(\vQ,\vQ')~+ f_2 \left( \vQ \cdot \vQ'~- {{\vec
p}\cdot \vQ
{}~{\vec p}\cdot \vQ' \over p_0^2}\right) ~+ f_3 {{\vec p}\cdot \vQ ~{\vec
p}\cdot \vQ' \over
p_0^2} \eqno(28)
$$
$$
\eqalignno{
f_1 ~&= {i \vp \cdot \vQ \over (p_0 -\vp\cdot \vQ)}\cr
f_2~&= i\sigma {1\over (p_0-\vp\cdot\vQ -{\twthird}\sigma)} \left({
p_0+\vp\cdot\vQ \over
p_0-\vp\cdot\vQ}\right)\cr
f_3~&= -{\twthird}\sigma ~{p^2\over p_0^2} ~{1\over \left( p_0-\vp\cdot\vQ -
{\twthird}\sigma~
({p^2\over p_0^2}) \right)} &(29)\cr
\sigma ~&= {k\over 4\pi}{p_0 \over p^2}\cr}
$$
We have used the vacuum average. A thermal distribution for the soft modes can
be
taken into account by imposing periodicity in imaginary time.
The correlator (27) is useful, for instance, in calculating the scattering of a
gluon
in the quark-gluon plasma. It is what defines the gluonic structure functions
of the plasma.

We now turn to a slightly different topic.
The Hamiltonian
analysis and the equations of motion can be considerably simplified for the
longitudinal modes.
Eventhough the restriction to just the longitudinal modes is not physically
well motivated, the simplifications are interesting and could eventually be
incorporated in a more general framework. For this reason, we shall briefly
discuss the reduction to longitudinal modes.

The longitudinal modes can be parametrized in terms of a matrix field $U$ as
$$
A_i ~=~-\partial_iU~U^{-1} \eqno(30)
$$
$U$ is in general time-dependent and hence the gauge potential (30) is not a
pure
gauge. The magnetic field, however, is zero. The electric field can be
parametrized in terms of a field $\Pi$ as $E_i = -U(\partial_i\Pi )U^{-1}$.
We can also define currents
$$
j_{\pm}~=~ U^{-1} J_{\pm}U \eqno(31)
$$
$j_{\pm}$ and $\Pi$ are gauge invariant degrees of freedom; the field $\Pi$
is constrained
in terms of $j_{\pm}$ by the Gauss law. Since $J_{\pm}$ commute with
$A_i$, they obviously commute with $U$. Using this fact,
the operator algebra can be simplified. The nontrivial commutators become
$$
\eqalignno{
[~\partial_i \partial_i \Pi^a (x),\partial_j \partial_j\Pi^b(y) ~]~&=
 if^{abc}  \partial_k \partial_k\Pi^c(x)
\delta (\vx-\vy) &(32a)\cr
[~-\partial_i \partial_i\Pi^a(x), j^b_{\pm}(y,\vQ)~]~&=
-if^{abc}  j^c_{\pm}(x,\vQ)\delta(\vx-\vy)\cr
&~~~~{\pm}{ik\over 4\pi} \delta^{ab} Q_i \partial_i
 \delta(\vx-\vy)
&(32b)\cr
[~j^a_{\pm}(x,\vQ), j^b_{\pm}(y,\vQ')~]~&= if^{abc}  j^c_{\pm}(x,\vQ)
\delta(\vx-\vy) \delta(\vQ,\vQ')\cr
& ~~~~{\mp}{ik\over 4\pi}\delta^{ab} Q_i\partial_i
 \delta(\vx-\vy)\delta(\vQ,\vQ') &(32c)\cr}
$$

The Hamiltonian becomes
$$
{\cal H}=\int d^3x~\left[ {\half}(\partial_i\Pi^a \partial_i \Pi^a) ~+
{2\pi \over k}\int d\Omega (j^a_+ j^a_+ ~+~ j^a_- j^a_- ) \right] \eqno(33)
$$
The Hamiltonian does not involve the field $U$ directly and hence the
commutator of $\Pi^a$ with $U$, which is nontrivial, is not needed to analyze
the
dynamics of the longitudinal modes.

The Gauss law condition becomes
$$
-\partial_i \partial_i \Pi^a ~+~ \int d\Omega~(j^a_+~+~j^a_- ) ~=0 \eqno(34)
$$
The commutation rules (32) are consistent with imposing the Gauss law as a
strong
operator condition. We may thus take (34) as the definition of $\Pi^a$
and the Hamiltonian and the commutation rules (32c) define the
dynamics, with one constraint, viz. $j^a_+ (x-\vQ)- j^a_-(x,\vQ)=0$.

The Hamiltonian equations of motion are now
$$
\partial_0 C^a_1 ~-~ Q_i\partial_i C^a_0 ~+~f^{abc}C^b_0 C^c_1 ~=0 \eqno(35)
$$
where $ j^a_+ = {k\over 4\pi} C^a_1,~ C^a_0 = \Pi^a -C^a_1$.
This equation can also be written as
$$
(\partial_0 + Q_i\partial_i)C^a_1 ~+~f^{abc}\Pi^b C^c_1~= ~Q_i\partial_i\Pi^a
\eqno(36)
$$
We have a similar equation for $C'^a_1= {4\pi\over k}j^a_-$, obtained
from (36) by $\vQ \rightarrow -\vQ$.
The Gauss law is
$$
-\partial_i \partial_i \Pi^a ~+{k\over 4\pi}\int d\Omega~
(C^a_1 +C'^a_1)~=0 \eqno(37)
$$
The solution of these equations, if we neglect the term involving the structure
constants $f^{abc}$, are the Abelian longitudinal plasma waves. Equation (36)
can
thus be used to systematically improve on the Abelian approximation.
\vskip .2in
\noindent{\bf Appendix A}
\vskip .2in
As is well known, WZNW-theories for unitary matrices lead to a quantization
requirement on the constant $k$ multiplying the action. This arises from the
single-valuedness of the transition amplitudes or wavefunctions.
For our case, eventhough the auxiliary field $G$ is a unitary matrix,
there is no such requirement. The Wess-Zumino term which leads to the
quantization
requirement can be written, for a fixed choice of $Q_i$, as
$$
{\cal S}_{WZ} ={1\over 12\pi}
\int d^2x^T \int_{M^3} {\rm Tr}(dG~G^{-1})^3 \eqno(A1)
$$
where $M^3$ is a space whose boundary is the two-dimensional world of
the lightcone coordinates $x^{\pm}$. ${\cal S}_{WZ}$ can be thought of
as being obtained by integrating $\delta {\cal S}_{WZ}$ with
the extra coordinate of $M^3$, say $s$,
parametrizing the path of integration.
Physical results are independent of this path. The difference for
two extensions of the field $G$ into the extra dimension corresponding to
$s$ is given by
$$
Q~={1 \over 12\pi} \int_{S^3} {\rm Tr}(dG~G^{-1})^3 \eqno(A2)
$$
This is for the intrinsic two-dimensional problem, ignoring the
transverse coordinates. $S^3$ is the three-dimensional sphere.
$Q$ is an integer corresponding to the winding number of the map
$G$ over the three-sphere. (The nontriviality of (A2) has to do
with the fact that ${\rm Tr}(dG~G^{-1})^3$ is a closed but nonexact
differential form.)
The single-valuedness of $e^{ikS_{WZ}}$ then
requires $k$ to be an integer.

For our case, we have the two transverse coordinates as well;
we can write equation (A1) as
$$
{\cal S}_{WZ}= {1\over 12\pi}\int_{M^5} {\rm Tr}(dG~G^{-1})^3 ~ \omega
\eqno(A3)
$$
where $\omega = \epsilon_{ijk}Q^i ~dx^j~dx^k$. There is no restriction for
the exterior derivatives in (A3) to be with respect to $x^{\pm}$ or $s$; this
is enforced by the choice of $\omega$. Notice that eventhough ${\rm Tr}(dG~
G^{-1})^3$ is not exact, the differential form in (A3) will be, if $\omega$ is
exact. $\omega$ can be formally written as $d\rho$, but the one-form $\rho$ is
linear in the coordinates and so we cannot use this unless we impose
some stringent fall-off conditions on the fields $G$. However, when we
take the difference of two extensions of $G$ into $M^5$, we get the quantity
on the right hand side of (A3) integrated over a five-sphere. $\rho$ exists
with
no singularity and the integrand is exact. There is no nonzero
winding number and no quantization of $k$.
\vskip .2in
\noindent{\bf Appendix B}
\vskip .2in
The Hamiltonian for the effective action was obtained
in ref.[5] by taking
the derivative of the action with respect to the final value
of time in a discretized version. This is a standard result, but
does not seem to be well appreciated. A short argument justifying
the calculation will be sketched here.

We shall use the example of a scalar field to illustrate the calculation.
Consider the Lagrangian
$$
L= {\half} \left( {{\partial \phi} \over {\partial t}}\right)^2
-V(\phi ) \eqno(B1)
$$
where $V(\phi )$ includes terms which do not depend on
time-derivatives of $\phi $.
The Hamiltonian, by definition, is
the generator of reparametrizations of the time-variable $t$.
Let us consider therefore the infinitesimal transformation
$t \rightarrow {\tilde t} =t+\epsilon (t)$, where $\epsilon (t)$ is a function
of $t$.  We then have $d{\tilde t} = dt(1+\epsilon'),
{}~ \partial_{\tilde t}=(1-\epsilon')\partial_t$, where the prime denotes
differentiation with respect to $t$. For the variation of the
action $S=\int dt~L$, we then find
$$
\delta S= -\int dt~ \epsilon'\int d^3x~
 \left[ {\half} \left({{\partial \phi}\over
{\partial t}}\right)^2 +V(\phi )\right]~+~
\int {\cal E}~ \delta \phi\eqno(B2)
$$
$\delta \phi$ is the change induced in $\phi$ and ${\cal E}$ is the variational
derivative. (Setting ${\cal E}$ to zero will give the equations of motion.)
The final limit of the time-integration is not changed in this variation, only
the coordinate label for it is varied.
{}From equation (B2) we can identify the Hamiltonian as
$$
H=\int d^3x~ \left[{\half} \left({{\partial \phi}\over
{\partial t}}\right)^2 +V(\phi )\right] \eqno(B3)
$$
If one imposes the equation of motion, viz.
${\delta S\over \delta \phi}=0$, one can prove the conservation of
$H$. However, to identify the expression for $H$ we do not impose
the equations of motion; we just use the separation of $\delta S$
in the form given in (B2).

We can recast the above result as
follows. Let us write the integral over time by
discretizing the time-interval, with $t_N$ being the final
time-label.
Then $\epsilon'(t_n)=
{{\epsilon_n -\epsilon_{n-1}}\over {t_n-t_{n-1}}}$, etc.
The action can be written as
$$
S= \int d^3x\left[ {\half} \sum {{(\phi_{n-i}-\phi_{n-i-1})^2}\over
{(t_{n-i} -t_{n-i-1})}}~- \bigl(
V(\phi)\bigr)
_{n-i}(t_{n-i}-t_{n-i-1})\right] \eqno(B4)
$$
The expression (B2) for $\delta S$ can be written as
$$
\delta S = - \left[ (\epsilon_N -\epsilon_{N-1})H(t_N )+...\right]
{}~+\int {\cal E}~\delta \phi \eqno(B5)
$$
We see that $H$ can also be identified as the coefficient of
$\epsilon_N$ in the first set of terms. However, $\epsilon_N$ is
the shift in $t_N$; thus
the coefficient of $\epsilon_N$ is the derivative of the action
with respect to
the final time-label $t_N$. However, in taking this derivative
we should not vary the fields as functions of time so as to
avoid contributions from the second set of terms, viz. those
proportional to ${\cal E}$.
This qualification means that we can use
expression (B4) for the action, but interpret $\phi_{n-i}$ as
independent variables, not as $\phi(t_{n-i})$.
Further, although we change $t_N$, the number of segments of time interval is
fixed.
The limits of the summation are unchanged, reflecting the fact that the
variation in
(B2) only changes the coordinate labels.
Applied to the effective action, the above argument gives the Hamiltonian (12),
i.e.
the result in ref.[5].
\vskip .2in
\noindent{\bf Appendix C}
\vskip .2in
In this appendix, we discuss the canonical derivation of
the operator algebra.

The general method is as follows. Given an action $S$ depending
on a set of fields $\varphi_r,~r=1,2,...n$, we consider
its variation under a change of fields. This is of the form
$$
\delta S= \int d^4x ~{\cal E}_r \delta\varphi_r~+~
\oint K_\mu ~d\sigma^\mu \eqno(C1)
$$
where ${\cal E}_r$ is the variational derivative.
The second term is the surface
contribution, with $K^\mu$ being linear in the
variations $\delta\varphi_r$. If we consider the time-coordinate as the
evolution
parameter and impose suitable boundary conditions at spatial
boundary (viz. that $K_i$ vanishes there), we can write the
surface term as the difference of $\Theta$'s at the initial and
final time-slices, where
$$
\Theta = \int d^3x~K_0 \eqno(C2)
$$
$\Theta$ is the canonical one-form. The variation of $\Theta$,
antisymmetrized in the variations, (i.e. the exterior derivative
on the field space ) gives the symplectic structure $\omega$.
(This is an old result going back to nineteenth century
analytical mechanics; in the context of quantum field theory,
Schwinger's action principle is closely related to this.
For some modern references, see [10]. Also,
an analogous method can be used even if the initial and final
data hypersurfaces do not correspond to constant time-slices.)
There are ambiguities in identifying $\Theta$ as above,
since we can add total divergences to the Lagrangian; $\omega$
has no such ambiguity. (The changes in $\Theta$ given by such
total divergences are, of course, the canonical transformations.)

Given $\omega$, the Poisson brackets for the canonical coordinates
are given
by the inverse to $\omega_{ij}$, viewed as an antisymmetric
matrix, where
$\omega ={\half }\omega_{ij}\delta \sigma^i~ \delta \sigma^j$,
and
$\sigma^i$ denote the canonical variables (like the fields or
their time-derivatives).
The commutators are then $i$ times the Poisson brackets.
For the present problem, we need commutators for currents, rather than
the canonical coordinates, and for this purpose it is algebraically
simpler to use an equivalent definition.
Given $\omega$, a vector field $\xi_A$ is said to be Hamiltonian if
$\xi_A\rfloor \omega =-\delta A$, for some function $A$;
alternatively, given any function $A$, we can associate a vector field
to it by this condition. The notation is as follows. If we write
$\omega$ in terms of its antisymmetric components $\omega_{ij}$
and $\xi = \xi^i {\delta \over \delta \sigma^i}$, then
$$
\xi \rfloor \omega ~=~ \xi^i \omega_{ij}\delta \sigma^j \eqno(C3)
$$
The commutator of two functions $A$ and $B$ is given by
$$
[~A,B~]~= i(\xi_A\rfloor \xi_B\rfloor \omega ) \eqno(C4)
$$

This method can be applied to our problem.
The only technical complication is that the WZ-term requires an
extension of the fields into an extra dimension.
We can take the extended space $M^3$ as $M^2\times [0,1]$, $s=1$
being real spacetime. The variations of the fields at $s=0$ can
be taken to be zero. From the variation of the effective action and after
identifying
$\Theta$, the symplectic structure
can be obtained as
$$
\eqalignno{
\omega ~= \int d^3x~\delta E^a_i \delta A^a_i ~+{k\over 4\pi}
\int d^3x~d\Omega~&
{\rm Tr}\bigl[ v ~\delta L_0 + v^2(L_0+Q_iL_i) \cr
&~+vQ_i\delta A_i
+G^{-1}v G~Q_i\delta A_i\bigr] &(C5)\cr}
$$
where $v= \delta G~G^{-1},~L_\mu =\partial_\mu G~G^{-1},~R_\mu =
G^{-1}\partial_\mu G$. The Hamiltonian vector fields are
$$
\xi_A ~= -{\delta \over \delta E^a_i},~~~~~~\xi_E ~=
{\delta \over \delta A^a_i}- \int d\Omega~ Q_i \left(
{\delta \over \delta L^a_0}+{\delta \over \delta R^a_0}\right)
\eqno(C6)
$$
For the variation $\delta G=\theta G$, the vector field is
$$
\xi_\theta = V(\theta )+\int d^3x~d\Omega~{k\over 8\pi}Q_i
\theta^b (\delta^{ba}-2M^{ba}){\delta \over \delta E^a_i} \eqno(C7)
$$
where $M^{ab}= {\rm Tr}[(-it^a)G(-it^b)G^{-1}]$ and
$V(\theta )$ replaces $v$ by $\theta$ and $\delta L_0$ by $[\theta,
L_0+Q_iL_i ]$. In other words,
$$
V(\theta )\rfloor \omega = {k\over 4\pi} {\rm Tr} \left[ \theta
(\delta L_0 +Q_i\delta A_i +G Q_i\delta A_i~G^{-1})\right]
\eqno(C8)
$$
We can calculate the various commutators using these vector fields
and the formula (C4). The result is the algebra (17) in the text.
\vskip .2in
\noindent {\bf References}
\vskip .2in
\item{[1]} R. Pisarski, {\it Physica} A 158, 246 (1989); {\it Phys.Rev.Lett.}
63, 1129 (1989); E. Braaten and R. Pisarski, {\it Phys.Rev.}
D 42, 2156 (1990); {\it Nucl.Phys.} B 337, 569 (1990); {\it ibid.} B 339,
310 (1990); {\it Phys.Rev.} D 45, 1827 (1992);
J. Frenkel and J.C. Taylor, {\it Nucl.Phys.} B 334, 199 (1990);
J.C. Taylor and S.M.H. Wong, {\it Nucl.Phys.} B 346, 115 (1990).
\vskip .1in
\item{[2]} R. Efraty and V.P. Nair, {\it Phys.Rev.Lett.} 68, 2891 (1992);
{\it Phys.Rev} D 47, 5601 (1993).
\vskip .1in
\item{[3]} R. Jackiw and V.P. Nair, {\it Phys.Rev.} D 48, 4991 (1993)
\vskip .1in
\item{[4]} J.P. Blaizot and E. Iancu, {\it Phys.Rev.Lett.} 70, 3376
(1993); Saclay preprint T93/064, to appear in {\it Nucl.Phys.} B.
\vskip .1in
\item{[5]} V.P. Nair, {\it Phys.Rev.} D 48, 3432 (1993).
\vskip .1in
\item{[6]} R. Jackiw, Q. Liu and C. Lucchesi, MIT preprint CTP\#2261
(1993);
J.P. Blaizot and E. Iancu, Saclay preprints T94/02,03, 013 (1994).
\vskip .1in
\item{[7]} S.P. Novikov, {\it Usp.Mat.Nauk} 37, 3 (1982); E. Witten,
{\it Commun.Math.Phys.} 92, 455 (1984).
\vskip .1in
\item{[8]} R.I. Nepomechie, {\it Phys.Rev.} D 33, 3670 (1986); D. Karabali,
H.J. Schnitzer and Z. Yang, {\it Phys.Lett.} 216 B, 307 (1989);
D. Karabali and H.J. Schnitzer, {\it Nucl.Phys.} B 329, 649 (1990);
K. Gawedzki and A. Kupianen, {\it Phys.Lett.} 215 B, 119 (1988);
{\it Nucl.Phys.} B 320, 649 (1989).
\vskip .1in
\item{[9]} see for example, P.V. Landshoff, {\it Nucl.Phys.} B {\it
Proc.Suppl.} 16, 597 (1990).
\vskip .1in
\item{[10]} V. Guillemin and S. Sternberg, {\it Symplectic Techniques
in Physics}, Cambridge University Press (1990); J. Schwinger,
{\it Phys.Rev.} 82, 914 (1951); C. Crnkovic and E. Witten, in
{\it Three Hundred Years of Gravitation}, S.W. Hawking and W. Israel (eds.),
Cambridge University Press (1987); G.J. Zuckerman, in
{\it Mathematical Aspects
of String Theory}, S.T. Yau (ed.), World Scientific (1987).
\end